\documentclass[aps,pra,reprint,superscriptaddress,twocolumn,showpacs,amsmath]{revtex4-1}

\usepackage{amssymb}
\usepackage{graphicx}
\usepackage{bm}
\usepackage{hyperref} 

\def\mywidtha{\the\dimexpr1\columnwidth}
\def\mywidthb{\the\dimexpr0.8\columnwidth}

\usepackage{enumitem}

\usepackage{epstopdf}
\usepackage{amsmath,amssymb}
\usepackage[percent]{overpic}
\usepackage{color}
\usepackage{hyperref}

\usepackage{fancyhdr}
\usepackage[cm]{fullpage}
\newcommand{\abs}[1]{\left| #1 \right|}

\begin{document}

\title{Nonlinear imaging in photonic lattices}

\author{Nikolaos K. Efremidis}
\email[Corresponding author: ]{nefrem@uoc.gr}
\affiliation{Department of Mathematics and Applied Mathematics, University of Crete, 70013 Heraklion, Crete, Greece}

\author{Mihalis Mparkas}
\affiliation{Department of Mathematics and Applied Mathematics, University of Crete, 70013 Heraklion, Crete, Greece}

\date{\today}

\begin{abstract}
We show that nonlinear imaging is possible in periodic waveguide configurations provided that we use two different segments of nonlinear media with opposite signs of the Kerr nonlinearity with, in general, no other restriction about their magnitudes. The second medium is used to implement effective ``reverse propagation''. A main ingredient in achieving nonlinear imaging is the control of the sign and the amplitude of the coupling coefficient. We numerically test our results in one and two dimensional square arrangement of waveguides. 
\end{abstract}

\maketitle

The study of systems with periodic modulation of the refractive index is related to a host of novel phenomena~\cite{chris-ol1988,eisen-prl1998,fleis-nature2003,chris-nature2003,leder-pr2008,garan-PR2012}. A main ingredient in connection to this work is the possibility to engineer their effective diffraction. In this respect, it has been shown both in the linear and the nonlinear regimes that diffraction in waveguide arrays can be reduced, canceled, or even reversed~\cite{eisen-prl2000,ablow-prl2001,perts-prl2002,locat-ol2005}. Diffraction can be controlled by engineering the coupling coefficient between adjacent waveguides. A main tool in this respect is the phenomenon known as coherent destruction of tunneling (CDT)~\cite{gross-prl1991}. Utilizing CDT can lead to the inhibition of tunneling in waveguide arrays~\cite{stali-oe2006,szame-prl2009,karta-ol2009} as well as negative coupling and refraction~\cite{zhang-ol2010,zhang-apb2011}. What is particularly interesting in CDT is that individual couplings can be separately engineered in a waveguide network. For example, by generating a diatomic lattice with alternating positive and negative couplings, an analogue of the 1D Dirac equation can be implemented~\cite{efrem-pra2010,zeune-prl2012}. 

In the linear limit there are several ways to achieve imaging in periodic lattices.
Self-imaging effects are possible by utilizing optical Bloch oscillations~\cite{pesch-ol1998,perts-prl1999,moran-prl1999-bloch} as well as the phenomenon of dynamic localization~\cite{dunla-prb1986,digna-prl2002,lenz-oc2003,longh-ol2005,longh-prl2006}. The image recovery at the output plane can also be achieved by inducing an intervening $\pi$-phase shift in the array~\cite{longh-ol2008,szame-apl2008} even in the presence of 
disorder~\cite{keil-ol2012}. 
An alternative approach to achieve imaging is proposed in~\cite{vicen-jo2014} using the flat band modes of a Kagome lattice. The nonlinearity of the system can also be utilized in generating arrays of stable solitons for image transmission~\cite{yang-ol2011}. 

In the above cases, where linear recovery of the image is possible, the effect of Kerr nonlinearity is going to be detrimental. In bulk media a technique that is used in the nonlinear regime is that of reverse propagation~\cite{chang-ol2003,barsi-ol2009,goy-pra2011}: The signal propagates inside the medium and the output is recorded. Then numerical reverse propagation is involved to recover the original image. Of course the same principle can be directly applied in the discrete case. However, this in not what we intend to do here. 
We propose the use of a second medium that effectively inverts the direction of propagation. This can be achieved if the second medium has the opposite signs in the nonlinearity and the diffraction, with the same ratio between them so that effectively $z\rightarrow-\sigma z$. In bulk media the aforementioned restrictions are making such a possibility quite difficult to implement. However, as we will see, in artificial discrete media such a prospect can be realized. 

In this work, we show that nonlinear imaging in periodic waveguide configurations is possible by using a second subsequent array with the opposite sign (and arbitrary magnitude) of the nonlinearity. This is achieved by diffraction engineering using CDT with either amplitude or spatial frequency modulation. We verify our predictions via direct numerical simulations in both 1D and 2D configurations of waveguides. We have identified two different regimes in the dynamics; In the first one the initial wave is always recovered at the output, whereas in the second there is a certain power threshold above which nonlinear instabilities start to take place. It is important to note that independently of the sign of the Kerr nonlinearity we can always switch between these two regimes by control of the sign of the coupling coefficient. Thus, nonlinear imaging can be implemented for both self-focusing (SF) and self-defocusing (SD) nonlinearities.

Let us consider the coupled-mode theory (CMT) dynamics of an optical wave propagating in a periodic lattice~\cite{chris-nature2003,leder-pr2008}
\begin{equation}
i\dot{c}_{n}+F_nc_{n}+K(c_{n+1}+c_{n-1})+\Gamma\abs{c_{n}(z)}^2c_{n}(z)=0
\label{eq:cmt}
\end{equation}
where $c_n$ is the optical field of the $n$th waveguide, $\dot{c}_n=dc_n/dz$ and $z$ is the propagation direction. In Eq.~(\ref{eq:cmt}) $K$ is the coupling constant between two adjacent waveguides, $F_n$ is a detuning in the propagation constant which depends on the waveguide number, and $\Gamma$ is the effective Kerr nonlinear coefficient. 

\begin{figure}[tb]
	\centering
	\includegraphics[width=\mywidtha]{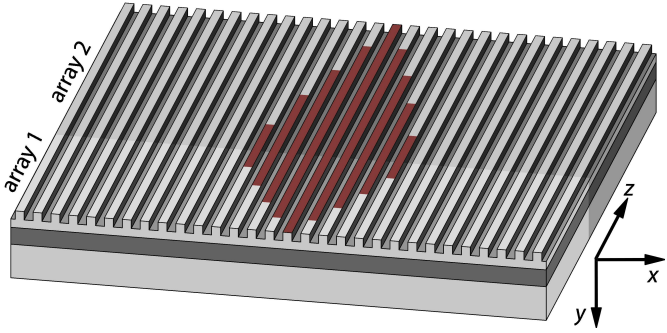}
	\caption{Schematic illustration of the proposed scheme. Two different waveguide configurations with opposite signs of the Kerr nonlinearity are utilized for the image reconstruction at the output.}
\label{f:goal}
\end{figure}
 Assume an optical wave of arbitrary shape propagating in a uniform waveguide lattice $c_n(z)$ ($n\in\mathbb Z$, $z\in[0,z^{(1)}]$), under the action of discrete diffraction and Kerr nonlinearity. Our goal is to utilize a second consequent array in such a way that the initial waveform is recovered at the output $[c_n(z^{(2)})=c_n(0)]$ (except perhaps from an additional constant phase) as it is schematically shown in Fig.~\ref{f:goal}.
Note that, if at $z=z^{(1)}$ we rescale the propagation distance as $z\rightarrow-\sigma z$ (with $\sigma>0$) then at $z^{(2)}=z^{(1)}(1+1/\sigma)$ (we added the absolute values of the two distances) the initial waveform is recovered. In our case, rather than changing the arrow of $z$ we change the effective coupling and nonlinear coefficients. Specifically, if $K=\kappa^{(j)}$ and $\gamma=\gamma^{(j)}$ $j=1,2$ inside the two media then we require that 
\begin{equation}
\kappa^{(2)}/\kappa^{(1)}=
\gamma^{(2)}/\gamma^{(1)}=-\sigma. 
\label{eq:condition}
\end{equation}
For implementation purposes, it is important that we do not make any assumptions about the magnitude of the nonlinearity of the two media and only constraint their sign to be opposite. Thus we obtain $-\sigma=\gamma^{(2)}/\gamma^{(1)}$. 
Note that the sign of the nonlinearity can be tuned by controlling the bias in photosensitive crystals in terms of the photoelectric effect~\cite{efrem-pre2002pr}.
On the other hand, the effective coupling coefficient can be controlled in periodically $z$-dependent waveguides~\cite{efrem-pra2010,zhang-ol2010,zhang-apb2011,zeune-prl2012}. 
The next step, described below, is to engineer the coupling coefficients so that they have the same ratio ($-\sigma$).

We assume that the propagation constants $F_n(z)$ are periodically modulated as
\begin{equation}
F_n(z) = \tfrac{(-1)^nA(z)}2\cos\theta(z),\quad\theta(z)=\alpha(z) z + \xi(z)
\label{eq:Fn}
\end{equation}
where the functions $A$, $\alpha$, and $\xi$ are piecewise constant 
$
G  =G^{(1)}+H(z-z^{(1)})(G^{(2)}-G^{(1)}), 
$
with $G=\{A,\alpha,\xi\}$, and $H$ is the Heaviside step function. 
Such a functional dependence can be implemented in different settings. For example, in the case of optically induced lattices~\cite{efrem-pre2002pr}, it can be realized by the interference of plane waves with different propagation constants~\cite{zhang-ol2010}. Furthermore, similar modulations have been observed in fs written curved waveguides~\cite{zeune-prl2012}. By applying the transformation 
$
c_n\rightarrow c_ne^{i(-1)^nA\sin\theta/(2\alpha)}
$
and assuming that the oscillating frequency is large enough we asymptotically obtain the effective coupling coefficient between adjacent waveguides 
\begin{align}
\kappa^{(1)} & = KJ_0(A^{(1)}/\alpha^{(1)}),\quad 0<z<z^{(1)}, \label{eq:kappa1} \\
\kappa^{(2)}_n & = KJ_0(A^{(2)}/\alpha^{(2)}) e^{i\phi_n},\quad z^{(1)}<z<z^{(2)}
\label{eq:kappa2}
\end{align}
where $J_0$ is the Bessel function of zeroth order, $\phi_n=(-1)^n\phi$, and 
$
\phi = 
 (A^{(2)}/\alpha^{(2)})\sin(\alpha^{(2)}z^{(1)}+\xi^{(2)})
-(A^{(1)}/\alpha^{(1)})\sin(\alpha^{(1)}z^{(1)}+\xi^{(1)}).
$
In general the addition phase term $\phi_n$ can be detrimental in the recovery of the initial waveform. 
Note that an interesting case appears when $\phi=\pi$ that leads to an effective inversion of the sign of the coupling due to the accumulated phase. 
In all our simulations below the selected parameters result $\phi=0$ for  $\xi^{(j)}=0$ and, thus, we rely in changes in the argument of the Bessel function in order to modify both the amplitude and the sign of the coupling coefficient. 

Setting the phase $\phi=0$, the condition for the nonlinear image retrieval becomes
\begin{equation}
\sigma J_0(A^{(1)}/\alpha^{(1)})+
J_0(A^{(2)}/\alpha^{(2)}) = 0.
\label{eq:condition}
\end{equation}
In the equation above we separate two different cases (i) amplitude modulation, where $\alpha^{(1)}=\alpha^{(2)}=\alpha$, $A^{(1)}\neq A^{(2)}$ and (ii) frequency modulation where $\alpha^{(1)}\neq\alpha^{(2)}$, $A^{(1)} = A^{(2)}=A$. In our simulations we choose to utilize amplitude modulation.

\begin{figure}[t]
\centerline{
\includegraphics[width=\mywidthb]{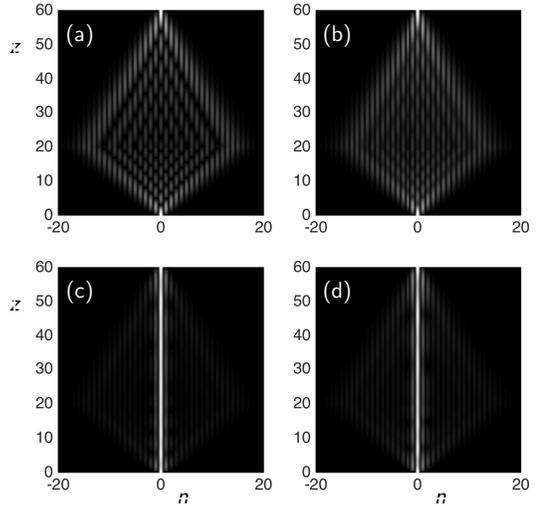}
}
\caption{Field amplitude dynamics in the case of single waveguide excitation $c_n(0) = \delta_{n,0}$ with $K=1$, and $\alpha=12$. 
In (a) the medium is linear whereas in (b) $\gamma^{(1)}=1$, $\gamma^{(2)}=-0.5$, (c)  $\gamma^{(1)}=2$, $\gamma^{(2)}=-1$, and (d)  $\gamma^{(1)}=-2$, $\gamma^{(2)}=1$. The interface distance is $z^{(1)}=20$ and thus imaging is achieved at $z^{(2)}=60$. \label{fig:1}}
\end{figure}

In all our figures below rather than changing the incident power we choose unitary maximum amplitudes and rescale the nonlinear coefficients of both media -- this procedure is mathematically equivalent. In addition, for demonstration purposes, we fix $\sigma=1/2$, $\kappa^{(1)}=0.4$, and $\kappa^{(2)}=-0.2$. Then from Eqs.~(\ref{eq:kappa1})-(\ref{eq:kappa2}) we obtain the lowest magnitude solutions: $A^{(1)}=20.35$, $A^{(2)}=34.04$.
Numerical simulations to the CMT equations are performed by utilizing a 4th order Runge-Kutta scheme.

In Fig~\ref{fig:1} the impulse response is chosen as an initial condition ($c_n(z=0)=\delta_{n,0}$). The linear case is presented In Fig~\ref{fig:1}(a). In~\ref{fig:1}(b) the first medium is SF but diffraction is still dominant. In~\ref{fig:1}(c) the nonlinear coefficients are doubled and the nonlinearity is now strong enough to support a discrete soliton that resides mainly in a single waveguide. Additional phenomena such as intensity oscillations and diffraction of a small portion of the beam are also observed. After the interface the diffracted part of the of the beam starts to converge and at the output the initial waveform is recovered. Very similar are the dynamics in Fig.~\ref{fig:1}(d) where the signs of the nonlinearity are inverted as compared to~\ref{fig:1}(c). Thus we have seen that in the case of single waveguide excitation imaging is achieved independently of the sign and the magnitude of the nonlinearity.  

\begin{figure}[t]
\centerline{
\includegraphics[width=0.9\columnwidth]{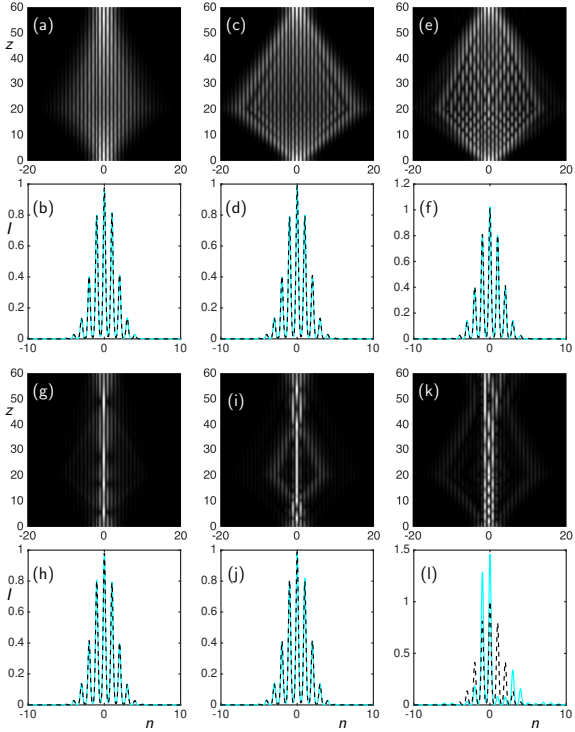}
}
\caption{Dynamics in the case of a Gaussian excitation of the waveguides
$u_n(z=0)=e^{-(n/3)^2}$. The parameters chosen are the same as those of Fig.~\ref{fig:1}. In (a)-(b) we see the linear case and in (c)-(f) the first medium is SD and the second SF with 
(c)-(d) $\gamma^{(1)}=-1$, $\gamma^{(2)}=0.5$ and 
(e)-(f) $\gamma^{(1)}=-2$, $\gamma^{(2)}=1$. 
In (g)-(l) the first medium is SF and the second SD with 
(g)-(h) $\gamma^{(1)}=0.5$, $\gamma^{(2)}=-0.25$, 
(i)-(j) $\gamma^{(1)}=1$, $\gamma^{(2)}=-0.5$, and
(k)-(l) $\gamma^{(1)}=2$, $\gamma^{(2)}=-1$. In the first and the third row the amplitude dynamics are shown whereas in the second and forth row a comparison of the intensity profiles of the final state (cyan solid) to the initial condition (black dashed) is depicted. 
\label{fig:2}}
\end{figure}
In multiwaveguide initial excitations the sign of $\mu^{(1)}=\gamma^{(1)}\kappa^{(1)}$ identifies two different regimes in the dynamics. Since we have selected $\kappa^{(1)}>0$ in our simulations, the relevant parameter is $\gamma^{(1)}$. In Fig.~\ref{fig:2} we depict the case of a Gaussian initial excitation of the waveguides. An additional $2\%$ noise multiplies the initial condition to break any possible symmetries that can suppress nonlinear instabilities. In Fig.~\ref{fig:2}(a)-(b) we see the linear case. In Fig.~\ref{fig:2}(c)-(f) the nonlinearity of the first medium is SD and this leads to increased diffraction. In the second SF medium the beam starts to refocus and at the output the initial waveform is recovered even in the case of strong nonlinearities [Fig.~\ref{fig:2}(e)-(f)]. The dynamics in the case where the first medium is SF exhibit several differences. When the nonlinearity of the first medium is low the beam diffracts and then refocuses in the second medium resulting to the recovery of the intensity pattern at the output. However, as the nonlinearity (or input power) increases discrete focusing and soliton formation starts to take place inside the first medium. Even in this case, as can be seen in Figs.~\ref{fig:2}(g)-(j), the beam manages to recover its original form at the output. However, there is a specific threshold that is determined by $\gamma^{(1)}P$, ($P$ is the total power) where nonlinear instabilities start to take place. As can be seen in Fig.~\ref{fig:2}(l) once this limit is reached imaging is no longer possible. Note that modulational instability is the main source for instabilities in such imaging configurations.

Note that the nonlinear instabilities occurring when the first medium is SF can be fully suppressed by changing the sign of the coupling $\kappa^{(1)}$ to negative (leading to $\mu^{(1)}<0$). Then the beam is going to diffract inside the first medium rather than focus. For example, the dynamics depicted in Figs.~\ref{fig:2}(k)-(l), once the sign of the couplings is reversed, become similar to those of the Figs.~\ref{fig:2}(e)-(f). An alternative approach with the same outcome requires applying a $\pi$ phase shift to the initial condition between adjacent waveguides [$(-1)^n$]. 

\begin{figure}
\centerline{
\includegraphics[width=0.9\columnwidth]{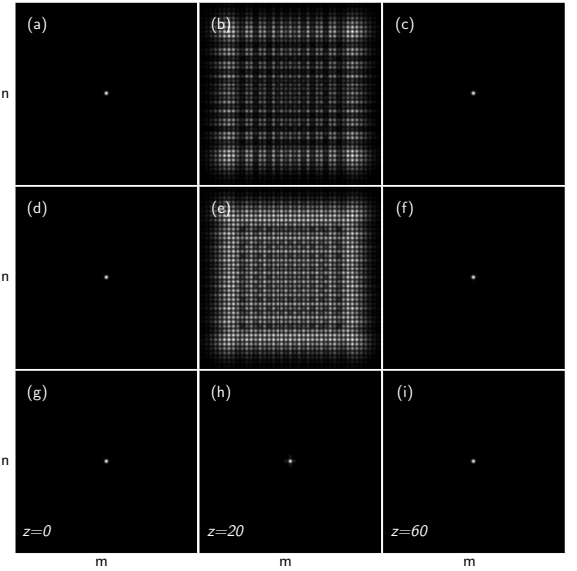}
}
\caption{Beam dynamics in the case of single waveguide excitation ($u_{m,n}(z=0)=\delta_{m,0}\delta_{n,0}$) at the input plane (left column, $z=0$), at the interface between the two media (middle column, $z=20$), and at output of the array (right column, $z=60$). The parameters are the same as those of Fig.~\ref{fig:1} with the difference that $\alpha=20$. In the first row the two media are linear, in the second row $\gamma^{(1)}=2$, $\gamma^{(2)}=-1$, and in the third row $\gamma^{(1)}=4$, $\gamma^{(2)}=-2$.\label{fig:2D}}
\end{figure}
The accuracy of the asymptotic formulas presented above increases as the spatial frequency increases.
This can be tested by numerically computing the Floquet quasienergies~\cite{zhang-ol2010,zhang-apb2011}. 
Furthermore, the accuracy of Eqs.~(\ref{eq:kappa1})-(\ref{eq:kappa2}) decreases with the dimensionality of the problem. 
This can be compensated by increasing the spatial frequency of the oscillations 
from $\alpha=12$ (in 1+1D) to $\alpha=20$ (in 2+1D).

The results of a single waveguide excitation in the case of a 2D square array are presented in Fig.~\ref{fig:2D}. In the first row the dynamics are linear. In the second and third rows the nonlinearity of the first medium is SF. In the second row the beam exhibits nonlinear diffraction, and then refocuses to the central waveguide at the observation plane. In the third row the nonlinearity is strong enough to support a 2D discrete soliton. The small portion of the power that diffracts refocuses at the observation plane.

\begin{figure}
\centerline{
\includegraphics[width=0.9\columnwidth]{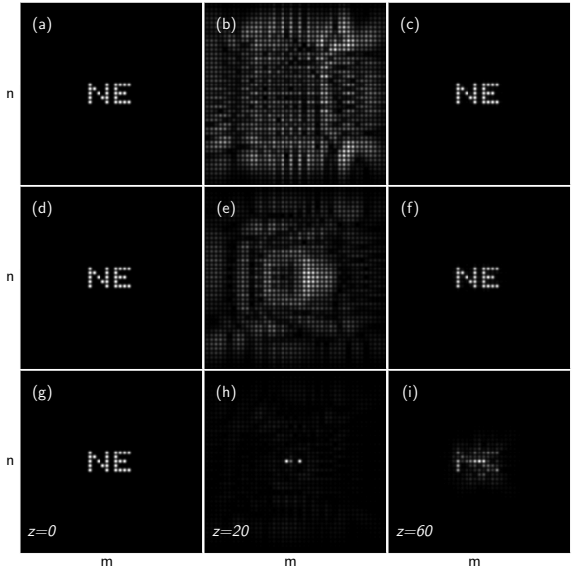}
}
\caption{Nonlinear imaging of the letter combination ``NE''. Each excited waveguide has unitary amplitude. The parameters are the same as those of Fig.~\ref{fig:2D}. In the three columns the results for $z=0,20,60$ are presented, respectively. In the first row the first medium is SD ($\gamma^{(1)}=-1$, $\gamma^{(2)}=0.5$), while in the second and third rows the first medium is SF ($\gamma^{(1)}=0.5$, $\gamma^{(2)}=-0.25$ and $\gamma^{(1)}=1$, $\gamma^{(2)}=-0.5$, respectively). 
\label{fig:NE}}
\end{figure}

Finally, in Fig.~\ref{fig:NE} we simulate the transmission of the letter combination ``NE''. In the case where the nonlinearity of the first medium is SF the beam diffracts until it enters the second medium. Then it refocuses, and at the observation plane the initial pattern is recovered (first row). In the second row the nonlinearity of the first medium is SF. We see in Fig.~\ref{fig:NE}(e) that the amount of diffraction is reduced as compared to Fig.~\ref{fig:NE}(b). Still, we are able to observe the revival of the initial pattern at the observation plane. In the third row we further increase the nonlinear coefficients. In this case, the total power carried by the beam is enough to generate discrete solitons along with surrounding radiation as shown in~\ref{fig:NE}(h). Such strong nonlinear effects lead to instabilities that prohibit the recovery of the initial waveform at the output. However, a careful inspection of Fig.~\ref{fig:NE}(i) shows that part of the initial intensity pattern is still regenerated. As in the 1+1D case, we are able to suppress these instabilities either by changing the signs of the couplings in the two media ($\mu^{(1)}<0$) or by imposing a phase $(-1)^{m+n}$ to the initial condition. Then the dynamics of the third row are transformed to those of the first row. 

In conclusion, we have shown that nonlinear imaging is possible in waveguide arrays provided that a second segment of an engineered array with opposite sign of the nonlinearity is used. 
We have presented numerical examples of nonlinear imaging in 1D and 2D arrangements of waveguides. We identified two different regimes according to the sign of $\mu^{(1)}=\gamma^{(1)}\kappa^{(1)}$. If $\mu^{(1)}$ is negative the image is always recovered, while if it is positive a certain nonlinear threshold exists above which nonlinear instabilities start to take place. Since $\mu^{(1)}$ does not depend solely on the sign of the nonlinearity ($\gamma^{(1)}$), we are able to always achieve stable dynamics by selecting the appropriate sign in the coupling coefficient ($\kappa^{(1)}$). Applications of such a configuration include nonlinear image transmission, multi-port couplers, and routing devices.

Funding: Erasmus Mundus NANOPHI Project (2013- 5659/002-001)

\newcommand{\noopsort[1]}{} \newcommand{\singleletter}[1]{#1}

\end{document}